\documentclass[preprint,aps,unsortedaddress,noshowpacs,a4paper,12pt]{revtex4}
\usepackage{graphicx}

\begin{document}

\title{A gyrofluid description of Alfv\'{e}nic turbulence and its parallel
electric field}
\author{N.H. Bian and E.P. Kontar}
\affiliation{Department of Physics and Astronomy, The University of Glasgow, G12 8QQ, United Kingdom}

\begin{abstract}
Anisotropic Alfv\'{e}nic fluctuations with $k_{\parallel}/k_{\perp}\ll 1$ remain at frequencies much smaller than the
ion cyclotron frequency in the presence of a strong background magnetic field. Based on the simplest truncation of the
electromagnetic gyrofluid equations in a homogeneous plasma, a model for the energy cascade produced by Alfv\'{e}nic turbulence is constructed,
which smoothly connect the large magnetohydrodynamics (MHD) scales and the small "kinetic" scales. Scaling relations are obtained for
the electromagnetic fluctuations, as a function of $k_{\perp}$ and $k_{\parallel}$. Moreover, a particular attention is paid to the
spectral structure of the parallel electric field which is produced by Alfv\'{e}nic turbulence. The reason is the potential implication
of this parallel electric field in turbulent acceleration and transport
of particles. For electromagnetic turbulence, this issue was raised some time ago in [A. Hasegawa, K. Mima,  J. Geophys. Res. {\bf 83} 1117 (1978)].
\end{abstract}

\maketitle

\section{Introduction}

At large wave-numbers in the direction perpendicular
to the ambient magnetic field, Alfv\'{e}n waves produce a significant compression of the plasma
which results in the creation of a parallel electric field via
the thermo-electric effect.
The parallel electric field associated with
a spectrum of kinetic Alfv\'{e}n waves leads to wave-particle interactions that can cause turbulent transport
phenomena. This was pointed out already some time ago in Reference [1], in the context of the magnetospheric plasma.
Starting from the drift-kinetics for the electrons, the authors of Ref.[1] derived a quasi-linear diffusion equation
in both momentum and coordinate space, a so-called double-diffusion, showing the pivotal role played by the parallel electric field spectrum.
In fact, their result is valid for any kind of low-frequency electromagnetic turbulence, but
they used the properties of the kinetic Alfv\'{e}n wave to give order of magnitude
estimates of various transport coefficients (density, momentum) and the electrons heating rate.
The solar wind is a natural laboratory plasma where the spectral properties of electromagnetic turbulence can be probed
with good accuracy down to scales below the ion Larmor radius. Turbulent Alfv\'{e}nic fluctuations observed in the solar wind have
a solar origin. Hence, beside their intrinsic interests, solar wind measurements can also be used as a diagnostic tool to infer the properties
of the turbulence closer to the sun, in the corona, where the plasma conditions are however different.
For low-frequency Alfv\'{e}nic fluctuations, the spectral energy density of the parallel electric field remains
a small fraction of the electro-magnetic energy density. Hence, it is hardly a measurable quantity.
However, since the parallel electric field is the actual force which mediates the wave-particle
interaction, it is interesting to investigate its spectral structure in this range of spatio-temporal
scales. This task is motivated by its potential relevance for turbulent acceleration and transport
and is facilitated by more recent advances in the understanding of Alfv\'{e}nic turbulence, mainly from
theory and observations of the solar wind.

The starting point is a minimal fluid model of the non-linear dynamics of low-frequency kinetic
Alfv\'{e}n waves, also capable
of capturing the effect associated with the thermal Larmor radius of the ions, $\rho_{i}=v_{T_{i}}/\omega_{ci}$ with $v_{T_{i}}=\sqrt{T_{i}/m_{i}}$
and $\omega_{ci}=eB_{0}/m_{i}c$. This model can be obtained as a truncation of the electromagnetic
gyrofluid equations in a constant background magnetic field, by assuming a constant temperature for both
the ions and the electrons, a constant background ion density and discarding all but the lowest two parallel moments
of the electron kinetics (see [2-5] and references therein).
These gyrofluid equations are moments of the gyrokinetic equations.
Gyrokinetics owe its name from being based on an averaging of the kinetic and Maxwell equations over the gyromotion of the particles and, hence, it applies in the limit of frequencies small compared to the ion cyclotron frequency, $\omega\ll \omega_{ci}$.
Extensive reviews on gyrokinetics can be found
in Refs.[6-8]. Gyrokinetics and their fluid moments are valid in the limit of small Larmor radius, $\rho_{i}\ll L$, $L$
being the macroscopic length scale of the plasma, and small anisotropic fluctuations, in the sense that $k_{\parallel}/k_{\perp}\sim \delta B/ B_{0}\ll 1$, $B_{0}$ being the magnitude of the background magnetic field. As for reduced-MHD\cite{KP,St,H1}, these ordering imply
pressure balance in the direction perpendicular to the background magnetic field, such that the fast mode is ordered out.
Moreover, because the gyroaverage procedure eliminates the cyclotron resonance, the only type of wave-particle interaction that remains
possible is through the Landau resonance between the particles and the parallel electric force or the magnetic mirror one.
On other central assumption of gyrokinetics, and hence of its fluid counterpart, is the small deviation of the distribution functions of the particles from a background distribution which is here a Maxwellian distribution.

The electromagnetic gyrofluid equations are presented first in Section II. Then they are used, in Section III, to construct a model of the energy cascade of Alfv\'{e}nic turbulence from the large MHD scales down to the small kinetic scales. The spectral structure of the parallel electric field produced by Alfv\'{e}nic turbulence is studied in Section IV. Finally, possible implications of the existence of this parallel electric field
are discussed in Section V.

\section{The electromagnetic gyrofluid model}

The gyrofluid model considered in this work involves the electron density $n_{e}$,
the magnetic flux function $\psi=-A_{z}$, where $\mathbf{A}$ is the vector potential, and the
electrostatic potential $\phi$, $z$ being the coordinate along the background field $\mathbf{B}_{0}$\cite{WM, SH, LH, G}.
Adopting an MHD normalization (see the Appendix), these equations are
\begin{equation}
\partial_{t}n_{e}+[\phi, n_{e}]-\nabla_{\parallel}J=0,
\end{equation}
\begin{equation}
\partial_{t}\psi+\nabla_{\parallel}(\rho_{s}^{2}n_{e}-\phi)=0.
\end{equation}
The Poisson bracket is defined as $[f,g]=\mathbf{z}.\nabla f\times \nabla g$,
where $\mathbf{z}$ is the unit vector along $\mathbf{B}_{0}$, and $\nabla_{\parallel}f=\partial_{z}f+[\psi,f]$ for any fields $f$
and $g$. The quantity $J=\nabla_{\perp}^{2}\psi$ is the parallel current
density and $\rho_{s}=c_{s}/\omega_{ci}L$ is the normalized ion sound Larmor radius
with $c_{s}=\sqrt{T_{e}/m_{i}}$.
This system is closed by the gyrokinetic Poisson equation,
\begin{equation}
n_{e}=\frac{(\Gamma_{0}-1)}{\rho_{i}^{2}}\phi,
\end{equation}
where $\Gamma_{0}$ is an integral operator which describes the average of the electrostatic potential over a ring
of Larmor radius $\rho_{i}$. In Fourier space, $\Gamma_{0}(b)$  simply becomes
\begin{equation}
\Gamma_{0}(b)=e^{-b}I_{0}(b),
\end{equation}
where $b=\rho_{i}^{2} k_{\perp}^{2}$ and $I_{0}$ is the modified Bessel function of the first kind.
The above equations conserve the total energy given by
\begin{equation}
E=\int d\mathbf{x} [(\nabla_{\perp}\psi)^{2}+\rho_{s}^{2}n_{e}^{2}+\phi(1-\Gamma_{0})\phi/\rho_{i}^{2}].
\end{equation}
The model is valid for $v_{Ti}^{2}/v_{A}^{2}\ll 1$, and moreover, since the effects associated with the electron inertia are not taken
into account, it means that the conditions $v^{2}_{Te}/v_{A}^{2}\gg 1$ and $k_{\perp}d_{e}\ll 1$ must also be satisfied, $v_{A}$ being the Alfv\'{e}n
speed and $d_{e}$ the electron skin depth.
In other words, the present model only describes the "kinetic" regime of the dispersive Alfv\'{e}n wave, not its "inertial" regime.

The following comments are due. When a Pade approximant
of the operator $\Gamma_{0}(b)$
is used, which gives :
\begin{equation}
\Gamma_{0}(b)-1\approx-\frac{b}{1+b},
\end{equation}
the Poisson equation is converted into:
\begin{equation}
(1-\rho_{i}^{2}\nabla_{\perp}^{2})n_{e}=\nabla_{\perp}^{2}\phi.
\end{equation}
The model (1)-(3) with the Pade approximation (7), also including electron inertial effects,
was studied in the context of collisionless magnetic reconnection in Ref.[5]. A well known effect of the
electron inertia is the breaking of the frozen-in condition for the magnetic field.

When $k_{\perp}\rho_{i}\ll 1$, relation (7) expresses the fact that the density is also equal to the plasma
vorticity $n_{e}=\nabla^{2}_{\perp}\phi$,
in which case Eqs.(1)-(2) become equivalent
to a low-$\beta$ Hall-MHD model (see Appendix and e.g. Ref.[12]).
Reduced Hall-MHD is an extension of the standard reduced MHD with
the Hall and the electron pressure effect accounted for in the Ohms's law. The parallel component of the latter, i.e. Eq.(2),
can also be written as
 \begin{equation}\label{om}
 E_{\parallel}=
-\rho_{s}^{2}\nabla_{\parallel}n_{e},
\end{equation}
which shows the existence of a parallel electric field that arises from electron pressure gradient along
the field lines. Electrons are here assumed to be isothermal. On the contrary, $E_{\parallel}=0$ is the standard reduced MHD Ohm's law.
Equation (\ref{om}) is a relation between the parallel electric field
and the compressive density fluctuation, the parallel derivative being taken along the total magnetic field comprising
the background plus its perpendicular perturbation.
Implicit in Hall-MHD is the assumption of cold ions, or at least
$\tau\equiv T_{i}/T_{e}\ll 1$. On the contrary, the gyrofluid model is valid for arbitrary value of $T_{i}/T_{e}$.
An other simplification can be reached in the limit $k_{\perp}\rho_{i}\gg 1$, i.e. $n_{e}=-\phi/\rho^{2}_{i}$,
in which case Eqs.(1)-(3) become similar to the reduced electron-MHD (see Appendix and Ref.[8]).

Assuming that all fields vary like $e^{i(\omega t-\mathbf{k}.\mathbf{x})}$, the linearized equations (1)-(3) for the Fourier
components are
\begin{equation}
\omega n_{e}-k_{\parallel}k_{\perp}^{2}\psi=0,
\end{equation}
\begin{equation}
\omega \psi+k_{\parallel}(\phi-\rho_{s}^{2}n_{e})=0,
\end{equation}
\begin{equation}
n_{e}=(\Gamma_{0}(b)-1)\phi/\rho_{i}^{2}.
\end{equation}
Hence, this system yields the dispersion relation :
\begin{equation}
\omega^{2}=k_{\parallel}^{2}\rho_{s}^{2}k_{\perp}^{2}
(1-\frac{\tau}{\Gamma_{0}(b)-1}),
\end{equation}
with $\Gamma_{0}(b)=e^{-b}I_{0}(b)$. The Pade approximation
for the operator $\Gamma_{0}(b)$  gives
\begin{equation}
\omega^{2}=k_{\parallel}^{2}[1+k_{\perp}^{2}(\rho_{s}^{2}+\rho_{i}^{2})],
\end{equation}
which is a standard expression for the frequency of the Alfv\'{e}n wave in the "kinetic" regime\cite{HM}.

Let us notice that the gyrofluid derivation of the the Alfv\'{e}n wave frequency gives the same result
as its kinetic counterpart, however the latter also provides the imaginary part, i.e. the coefficient
associated with Landau damping. In the model proposed below for the energy cascade produced
by Alfv\'{e}nic turbulence, any form of dissipation, including this collisionless one, is neglected at all except the smallest scales
where it is supposed to balance the energy injected at large scales.
For Alfv\'{e}nic perturbation, relations between the density, the potential
and the flux function are provided by the equations (9)-(11), the magnitude of the perpendicular magnetic field fluctuation is given
by $B_{\perp}=k_{\perp}\psi$,  the
magnitude of the perpendicular component of the electric field
is given by $E_{\perp}=k_{\perp}\phi$ and the parallel component by $E_{\parallel}=\rho_{s}^{2}k_{\parallel}n_{e}$.

\section{Anisotropic Alfv\'{e}nic turbulence}
For Alfv\'{e}n waves, we observe from the energy integral Eq.(5), that the total energy density per wave-number
in the perpendicular plane, i.e. the one dimensional energy spectrum denoted by $E_{k_{\perp}}$, is
\begin{equation}
E_{k_{\perp}}=k_{\perp}\psi^{2},
\end{equation}
which coincides with the magnetic energy spectrum
$E_{k_{\perp}}=\delta B_{\perp}^{2}/k_{\perp}$.

Locality of the non-linear interactions and constancy of the
energy flux are assumed, hence the energy cascade rate $\epsilon$ is
\begin{equation}\label{flux}
\epsilon=\frac{k_{\perp}E_{k_{\perp}}}{\tau_{NL}}.
\end{equation}
On the other hand, the non-linear time scale is given by
\begin{equation}\label{tunl}
\tau_{NL}=\frac{1}{k_{\perp}^{2}\Phi}
\end{equation}
with the generalized stream-function $\Phi=\phi-\rho_{s}^{2}n_{e}$. Therefore, the non-linear time scale can also be written
as a function of the magnetic perturbation, i.e.
\begin{equation}
\tau_{NL}=[\rho_{s}k_{\perp}^{2}(1+\frac{\tau}{1-\Gamma_{0}})^{1/2}k_{\perp}\psi]^{-1},
\end{equation}
with $k_{\perp}\psi=(E_{k_{\perp}}k_{\perp})^{1/2}$. From (\ref{flux}), the energy spectrum is then obtained :
\begin{equation}
E_{k_{\perp}}\propto\epsilon^{2/3}\rho_{s}^{-2/3}k_{\perp}^{-7/3}(1+\frac{\tau}{1-\Gamma_{0}})^{-1/3}.
\end{equation}
The Pade approximation for the latter expression gives
\begin{equation}
E_{k_{\perp}}\propto\epsilon^{2/3}k_{\perp}^{-5/3}[1+k_{\perp}^{2}(\rho_{s}^{2}+\rho_{i}^{2})]^{-1/3}.
\end{equation}
When $T_{i}\ll T_{e}$, then $E_{k_{\perp}}\propto\epsilon^{2/3}k_{\perp}^{-5/3}(1+k_{\perp}^{2}\rho_{s}^{2})^{-1/3}$
showing a breakpoint for $k_{\perp}\rho_{s}\sim 1$. Hence, this is an Hall-MHD result. On the contrary, when $T_{e}\ll T_{i}$, then $
E_{k_{\perp}}\propto\epsilon^{2/3}k_{\perp}^{-5/3}(1+k_{\perp}^{2}\rho_{i}^{2})^{-1/3}$
with breakpoint given by $k_{\perp}\rho_{i}\sim 1$. In general, the breakpoint occurs at $k_{\perp}\rho\sim 1$ with $\rho=(\rho_{s}^{2}+\rho_{i}^{2})^{1/2}$. This breakpoint separates the MHD range with $E_{k_{\perp}}\propto \epsilon^{2/3}k_{\perp}^{-5/3}$
and the dispersive range with $E_{k_{\perp}}\propto \epsilon^{2/3}\rho^{-2/3}k_{\perp}^{-7/3}$.

Moreover, we have the following scaling relations: $B_{\perp}\propto\epsilon^{1/3}k_{\perp}^{-1/3}[1+k_{\perp}^{2}(\rho_{s}^{2}+\rho_{i}^{2})]^{-1/6}$,
$n_{e}\propto\epsilon^{1/3}k_{\perp}^{2/3}[1+k_{\perp}^{2}(\rho_{s}^{2}+\rho_{i}^{2})]^{-2/3}$,
$E_{\perp}\propto\epsilon^{1/3}k_{\perp}^{-1/3}[1+k_{\perp}^{2}(\rho_{s}^{2}+\rho_{i}^{2})]^{-2/3}(1+k_{\perp}^{2}\rho_{i}^{2})$.
Notice that $n_{e}$ above refers to density
fluctuations that arise from the kinetic Alfv\'{e}n wave compression.

The one dimensional energy spectrum for the perpendicular electric field is therefore $E_{\perp}^{2}/k_{\perp}\propto
\epsilon^{2/3}k_{\perp}^{-5/3}(1+k_{\perp}^{2}\rho^{2})^{-4/3}(1+k_{\perp}^{2}\rho_{i}^{2})^{2}$. For $T_{i}\sim T_{e}$, the latter
relation is a good fit to some solar wind measurements \cite{Bale,S1}. It corresponds to a power index $-5/3$ in the MHD regime
when $k_{\perp}\rho_{i}\ll 1$ and $-1/3$ in the dispersive range when $k_{\perp}\rho_{i}\gg1$.
When $T_{i}\gg T_{e}$, $E_{\perp}\propto\epsilon^{1/3}k_{\perp}^{-1/3}[1+k_{\perp}^{2}\rho_{i}^{2}]^{1/3}$,
which has only one break-point at $k_{\perp}\rho_{i}\sim 1$.
Notice however that in a plasma with
$T_{e}\gg T_{i}$ then,
$E_{\perp}\propto\epsilon^{1/3}k_{\perp}^{-1/3}[1+k_{\perp}^{2}\rho_{s}^{2}]^{-2/3}(1+k_{\perp}^{2}\rho_{i}^{2})$,
which has two break-points at $k_{\perp}\rho_{s}\sim 1$ and at $k_{\perp}\rho_{i}\sim 1$.

A theory for strong anisotropic Alfv\'{e}n wave turbulence was developed by Goldreich-Sidrar\cite{GS}. According to this theory, the anisotropy
of the turbulence is fixed by the condition,
\begin{equation}
\omega\sim \tau_{NL}^{-1},
\end{equation}
 which is a balance between the linear and the non-linear dynamical time scales.
This leads to the following scale dependent anisotropy relation :
\begin{equation}\label{an}
k_{\parallel}(k_{\perp})\sim \epsilon^{1/3}k_{\perp}^{2/3}[1+k_{\perp}^{2}(\rho_{s}^{2}+\rho_{i}^{2})]^{-1/6}.
\end{equation}
This is also equivalent to the ordering relation
\begin{equation}
\delta B_{\perp}\sim k_{\parallel}/k_{\perp}\ll 1,
\end{equation}
with $\delta B_{\perp}(k_{\perp})$ given above.
We end this section by emphasizing that the scaling relations for the energy spectrum and anisotropy
in the dispersive scales \cite{CB,S1} are similar to the ones of EMHD turbulence\cite{BC,NG,CL1,CL2}

\section {Parallel electric field spectrum}
Alfv\'{e}nic turbulence involves electric field fluctuations $\mathbf{E}$
which possess a component parallel to the magnetic field $\mathbf{B}$.
We now discuss the spectral structure of the parallel electric field produced by Alfv\'{e}nic turbulence.
The magnitude of the parallel electric field is small compared to the perpendicular one, hence hardly measurable and
not expected to change the total electric field spectrum.
 Nevertheless, the parallel electric field can be important for
  particle acceleration and cross-field transport induced by Alfv\'{e}nic turbulence, or in fact, by any type
of electromagnetic turbulence. The physical reason is that the parallel electric field
can efficiently accelerate particles along
the magnetic field lines which are bent in the direction perpendicular to $\mathbf{B}_{0}$ due to the perturbation $\mathbf{B}_{\perp}$.
Hence the correlation between acceleration and cross-field transport, and the pivotal role played by $E_{\parallel}$.
The results of the previous sections provide the scaling of the parallel electric field fluctuation :
\begin{equation}
 E_{\parallel}\propto\epsilon^{1/3}\rho_{s}^{2}k_{\parallel}k_{\perp}^{2/3}[1+k_{\perp}^{2}(\rho_{s}^{2}+\rho_{i}^{2})]^{-2/3},
\end{equation}
together with Eq.(\ref{an}) for the relation $k_{\parallel}(k_{\perp})$.
It is therefore easy to verify that the magnitude of $E_{\parallel}$, as a function
of $k_{\parallel}$, scales like
\begin{equation}
E_{\parallel}(k_{\parallel})\propto \rho_{s}^{2}k_{\parallel}^{2},
\end{equation} when $k_{\parallel}\ll \epsilon^{1/3}\rho^{-2/3}$, and like
\begin{equation}
E_{\parallel}(k_{\parallel})\propto \epsilon k_{\parallel}^{-1}\frac{\rho_{s}^{2}}{\rho_{s}^{2}+\rho_{i}^{2}},
\end{equation}
 when $k_{\parallel}\gg \epsilon^{1/3}\rho^{-2/3}$.
Moreover, the magnitude of $E_{\parallel}$
reaches its maximum,
\begin{equation}
E_{\parallel}  \sim \epsilon^{2/3}\frac{\rho_{s}^{2}}{(\rho_{s}^{2}+\rho_{i}^{2})^{2/3}},
\end{equation}
at the breakpoint $k_{\parallel}\sim \epsilon^{1/3}\rho^{-2/3}$, which corresponds to
to $k_{\perp}\rho \sim 1 $, with $\rho=(\rho_{s}^{2}+\rho_{i}^{2})^{1/2}$, where the two branches merge smoothly.
It is clear that the condition $k_{\perp}\rho \gg 1$ does not have do be realized for $E_{\parallel}$ to be of significant value
because the latter reaches its maximum precisely at the boundary between the MHD and the "kinetic" scales.

\section{discussion and conclusion}

Based on a simple truncation of the gyrofluid equations, we constructed a model for the energy cascade produced by Alfv\'{e}nic turbulence
which smoothly connect the large magnetohydrodynamics (MHD) scales and the small "kinetic" scales. A similar
model, with an emphasize on the effect of linear collisionless dissipation was proposed in Ref.[20]. Scaling relations are obtained for
the electromagnetic fluctuations as a function of $k_{\perp}$ and $k_{\parallel}$ and particular attention is paid to the
spectral structure of the parallel electric field produced by Alfv\'{e}nic turbulence.
The reason is that wave-particle interactions through
the parallel electric force can produce anomalous transport in both velocity and coordinate space.
A first principle understanding of such turbulent process is provided by
quasi-linear theory applied to the gyro-averaged drift-kinetic equations. Within the framework of quasilinear theory, the resulting diffusion equation for, say the electron distribution function $f$ reads (see Ref.1)
\begin{equation}
\frac{\partial f}{\partial t}=\sum_{k}\nabla[(\frac{e}{m_{e}})^{2}\frac{\pi}{2}\delta (\omega-k_{\parallel}v_{\parallel})E_{\parallel}^{2}\nabla f],
\end{equation}
with the operator $\nabla$ defined as
\begin{equation}
\nabla =\nabla_{v_{\parallel}}+\frac{k_{\perp}v_{\parallel}}{\omega \omega_{ce}}\nabla_{x}.
\end{equation}
All quantities here are dimensional, $v_{\parallel}$ is the component of the electron velocity parallel to the magnetic field and $x$ is the coordinate transverse to $\mathbf{B}_{0}$. The same type of equation can be written for the ion distribution function. A major ingredient entering the quasilinear diffusion coefficients, is the spectrum of the parallel electric field, which has been studied above for strong and anisotropic Alfv\'{e}nic turbulence.
The parallel electric field produced by the turbulence results in Landau damping of
the turbulent field. The effect of any scale dependent damping process, can
be accounted for on the energy cascade by writing a transport equation through wavenumber space as follow (see Ref.[20]):
\begin{equation}
\frac{\partial E_{k_{\perp}}}{\partial t}+\frac{\partial}{\partial k_{\perp}}[\frac{k_{\perp}}{\tau_{NL}}E_{k_{\perp}}]=-2\gamma E_{k_{\perp}},
\end{equation}
with $\gamma(k_{\perp}, k_{\parallel})$, representing the damping rate. Recently, the effect of Landau damping on the energy spectrum was studied this way, based on linear gyrokinetic calculations\cite{H2} of the damping rate. However, the modification of the
distribution function due to wave-particle interactions was not addressed. A possible approach is the standard quasi-linear theory. Let us notice that, in order for the fluid treatment to be valid, the number of accelerated particles must remain small compared to the number of core particles.
Modeling of such a turbulent acceleration mechanism through the parallel electric field spectrum of Alfv\'{e}nic fluctuations, its impact on the particles distribution function and back-reaction on the turbulent energy cascade, is the subject of ongoing work.


\section{Appendix}
In this appendix we provide few additional details concerning the gyrofluid model.
We also discuss some of its limiting cases (RMHD, ERMHD)
and the corresponding model energy spectra.
The normalized equations (1)-(2) follow from the electron continuity and the parallel
momentum equations, which can be written as
\begin{equation}\label{c}
\partial_{t} \frac{n_{e}}{n_{0}}+\frac{c}{B_{0}}[\phi,\frac{n_{e}}{n_{0}}]=
\partial_{z}v_{ez}+\frac{1}{B_{0}}[A_{z},v_{ez}]
\end{equation}
\begin{equation}\label{d}
\partial_{t}A_{z}+\frac{c}{B_{0}}[\phi,A_{z}]=-c\partial _{z}\phi+
\frac{cT_{e}}{e}\partial_{z}\frac{n_{e}}{n_{0}}+\frac{cT_{e}}{eB_{0}}[\frac{n_{e}}{n_{0}},A_{z}]
\end{equation}
with $v_{ez}=(c/4 \pi n_{0})\nabla_{\perp}^{2}A_{z}$. This system is
closed by the gyrokinetic Poisson equation
\begin{equation}\label{e}
n_{e}=(\Gamma_{0}-1)\frac{e n_{0}}{T_{i}}\phi,
\end{equation}
Adopting the following "MHD" normalization,
\begin{equation}
(\widehat{t},\widehat{x},\widehat{\phi},\widehat{A}_{z},\widehat{n}_{e})=(\frac{t}{\tau_{A}},\frac{x}{L},\frac{\phi c}{Lv_{A}B_{0}},\frac{A_{z}}{LB_{0}},
\frac{n_{e}\omega_{ci}\tau_{A}}{n_{0}}),
\end{equation}
with $\tau_{A}=L/v_{A}$, then Eqs.(\ref{c})-(\ref{e}) become Eqs.(1)-(3). Hence, there are two non-dimensional
parameters in this gyrofluid model : say, $\rho_{s}$ and $\rho_{i}$.
When $k_{\perp}\rho_{i}\ll 1$, the Poisson equation gives $n_{e}=\nabla^{2}_{\perp}\phi$,
and therefore Eqs.(1)-(2) are equivalent
to the following low-$\beta$ Hall-MHD model :
\begin{equation}
\partial_{t}\nabla^{2}_{\perp}\phi+[\phi, \nabla^{2}_{\perp}\phi]-\nabla_{\parallel}J=0,
\end{equation}
\begin{equation}\label{Ohm}
\partial_{t}\psi+\nabla_{\parallel}(\rho^{2}_{s}\nabla^{2}_{\perp}\phi-\phi)=0.
\end{equation}
In the MHD approximation, $E_{\parallel}=0$ and therefore the standard RMHD
reads :
\begin{equation}
\partial_{t}\nabla^{2}_{\perp}\phi+[\phi, \nabla^{2}_{\perp}\phi]-\nabla_{\parallel}J=0,
\end{equation}
\begin{equation}
\partial_{t}\psi-\nabla_{\parallel}\phi=0,
\end{equation}
Notice that the RMHD system is free of any non-dimensional parameter under the present normalization.
The RMHD conserves the energy :
\begin{equation}
E=\int d\mathbf{x} [(\nabla_{\perp}\psi)^{2}+(\nabla_{\perp}\phi)^{2}].
\end{equation}
Linearizing the RMHD system provides the frequency of the Alfv\'{e}n wave $\omega=\pm k_{\parallel}$.
Moreover, since the relation $\phi=\pm \psi$ holds for Alfv\'{e}nic fluctuations, it follows from the energy integral that
the one-dimensional energy spectrum is
\begin{equation}
E_{k_{\perp}}=k_{\perp}\psi^{2},
\end{equation}
which coincides with the magnetic energy spectrum, i.e.
$E_{k_{\perp}}=\delta B_{\perp}^{2}/k_{\perp}$ since $\delta B_{\perp}=k_{\perp}\psi$.
Assuming locality of the non-linear interactions and constancy of the
energy flux, the energy cascade rate $\epsilon$ is
\begin{equation}\label{flux2}
\epsilon=\frac{k_{\perp}E_{k_{\perp}}}{\tau_{NL}}.
\end{equation}
On the other hand, the non-linear time scale is given by
\begin{equation}
\tau_{NL}=\frac{1}{k_{\perp}^{2}\phi}
\end{equation}
Hence, this time scale can also be written in term of the magnitude of the magnetic fluctuation,
\begin{equation}
\tau_{NL}=\frac{1}{k_{\perp}^{2}\psi},
\end{equation}
with $k_{\perp}\psi=(E_{k_{\perp}}k_{\perp})^{1/2}$. The expression
for the energy spectrum in the MHD regime is then obtained :
\begin{equation}
E_{k_{\perp}}\propto\epsilon^{2/3}k_{\perp}^{-5/3}.
\end{equation}

An other limit consists in taking $k_{\perp}\rho_{i}\gg 1$, in which case
the Poisson equation (3) gives $\phi=-\rho_{i}^{2}n_{e}$, and hence Eqs.(1)-(2) become
\begin{equation}
\partial_{t}\phi+\rho_{i}^{2}\nabla_{\parallel}J=0,
\end{equation}
\begin{equation}
\partial_{t}\psi-(\frac{\tau}{1+\tau})\nabla_{\parallel}\phi=0.
\end{equation}
An extension of this Electron-RMHD model, valid for a wider range of values of $v^{2}_{Ti}/v^{2}_{A}$, was derived and
studied previously in Ref.[8]. This ERMHD conserves the
energy :
\begin{equation}
E=\int d\mathbf{x} [(\nabla_{\perp}\psi)^{2}+(\frac{\tau+1}{\tau})\frac{\phi^{2}}{\rho_{i}^{2}}].
\end{equation}
Linearizing the ERMHD system provides the frequency of the dispersive kinetic Alfv\'{e}n wave $\omega=\pm k_{\parallel}k_{\perp}\sqrt{\tau/1+\tau}\rho_{i}$.
Moreover, since the relation $\sqrt{\tau/1+\tau}(\phi/\rho_{i})=\pm k_{\perp}\psi$ holds for kinetic Alfv\'{e}n waves, it is easily verified that
the one-dimensional energy spectrum is again the magnetic energy spectrum
$E_{k_{\perp}}=k_{\perp}\psi^{2}$. The non-linear decorrelation time scale is given by
\begin{equation}
\tau_{NL}=\frac{1+\tau}{\tau}\frac{1}{k_{\perp}^{2}\phi},
\end{equation}
(see Ref.[8]). This is indeed the $k_{\perp}\rho_{i}\gg1$ limit of the relation (16), since $n_{e}=-\phi/\rho_{i}^{2}$ also in this limit.
Therefore, the expression for the energy spectrum in the dispersive range is
\begin{equation}
E_{k_{\perp}}\propto \epsilon^{2/3}\rho_{i}^{-2/3}(\frac{1+\tau}{\tau})^{1/3}k_{\perp}^{-7/3}.
\end{equation}

\end{document}